\documentclass[12pt]{article}
\usepackage{amsmath,amssymb,epsfig}
\usepackage{color}
\input{colordvi.tex}

\makeatletter

\renewcommand\section{\@startsection {section}{1}{\z@}%
                                   {-3.5ex \@plus -1ex \@minus -.2ex}%
                                   {2.3ex \@plus.2ex}%
                                   {\normalfont\large\bfseries}}

\renewcommand\subsection{\@startsection{subsection}{2}{\z@}%
                                     {-3.25ex\@plus -1ex \@minus -.2ex}%
                                     {1.5ex \@plus .2ex}%
                                     {\normalfont\normalsize\bfseries}}

\newcount\hour \newcount\minute
\hour=\time \divide \hour by 60
\minute=\time
\def\now{%
\ifnum \hour<13
  \ifnum \hour=0 \advance \hour by 12 \number\hour:\else \number\hour:\fi%
     \ifnum \minute<10 0\fi%
     \number\minute%
\ A.M.%
\else \advance \hour by -12 \number\hour:%
  \ifnum \minute<10 0\fi%
  \number\minute%
  \ P.M.%
\fi%
}

\makeatother

\begin{document}

% format
\baselineskip=18pt  % a la harvmac
\numberwithin{equation}{section}  % make eq labels (sec.num)
\allowdisplaybreaks  % allow page breaks in displayed eqs

% print date, time and filename
%\pagestyle{myheadings}
%\markright{{\tt \jobname.tex} -- \today{} \now}

%%%%%%%%%%%%%%%%%%%%%%%%%%%%%%%%%%%%%%%%%%%
%%%        TITLE BEGINS HERE
%%%%%%%%%%%%%%%%%%%%%%%%%%%%%%%%%%%%%%%%%%%

%% ========== title (note version) begins here ==========
%
%\vspace*{-1cm}
%\begin{center}
% {\Large\bf Title of the Document}
%\end{center}
%\vspace*{-.5cm}
%
%% ========== title (note version) ends here ==========

%% ========== title (paper version, a la harvmac) begins here ==========

\thispagestyle{empty}

% Report number
\vspace*{-2cm}
\begin{flushright}

%{\tt arXiv:yymm.nnnn}\\
 IFUP-TH/2009-15\\
%IPMU-07-0003
\end{flushright}

% title, authors, affiliation
\vspace*{2.7cm}

\begin{center}
 {\Large {\bf {On Confinement Index}}}\\

\vspace*{2cm}
{ \bf Kenichi Konishi$^{\dagger}$~and Yutaka Ookouchi$^{\ddagger }$\\}
 \vspace*{1.0cm}
 $^\dagger$ {\it Department of Physics, ``E. Fermi'',  University of Pisa,   
  \& INFN, Sezione di Pisa,      \\    Largo Pontecorvo, 3,  Pisa  56127, Italy }\\
 $^{\ddagger}$
{\it Perimeter Institute for Theoretical Physics, Waterloo, Ontario N2T1E7, Canada}\\[1ex]
 \vspace*{0.8cm}
% {\tt foo@bar},
% {\tt sh{}i{}ge{}@t{}h{{}}{e{ory}}.c{}a{}l{{}}te{{}}ch.e{{}}d{}u} % can this avoid spam?
\end{center}

\vspace{1cm} \centerline{\bf Abstract} \vspace*{0.5cm}

The smallest integer $t$ for which the Wilson loop $W^{t}$ fails to exhibit area law is known as the confinement index of a given
field theory.   The confinement index provides us with subtle information on the vacuum properties of the system.
 We study the behavior of the Wilson and  't Hooft loops and compute the confinement index in a wide class of ${\cal N}=1$ supersymmetric gauge theories.  All possible electric and magnetic screenings are taken into account. The results found are consistent with 
 the  $\theta$ periodicity, and whenever such a check is available, with the factorization property of Seiberg-Witten curves.

\newpage
\setcounter{page}{1} % don't number title page

%% ========== title (paper version, a la harvmac) ends here ==========

%%%%%%%%%%%%%%%%%%%%%%%%%%%%%%%%%%%%%%%%%%%
%%%           TITLE ENDS HERE
%%%%%%%%%%%%%%%%%%%%%%%%%%%%%%%%%%%%%%%%%%%

%\tableofcontents
%\printindex

%%%%%%%%%%%%%%%%%%%%%%%%%%%%%%%%%%%%%%%%%%%
%%%        MAIN TEXT BEGINS HERE
%%%%%%%%%%%%%%%%%%%%%%%%%%%%%%%%%%%%%%%%%%%

%%%%%%%%%%%%%%%%%%%%%%%%%%%%%%%%%%%%%%%%%%%%%%%%%%%%%%%%%%%%%%%%%%
\section{Introduction}
%%%%%%%%%%%%%%%%%%%%%%%%%%%%%%%%%%%%%%%%%%%%%%%%%%%%%%%%%%%%%%%%%%

Confinement is one of the fascinating and long-standing problems of the elementary particle physics today.  A widely accepted mechanism for confinement in QCD is the dual Meissner effect caused by condensation of some field with a  magnetic charge.
A power of supersymmetry sometimes gives us a quantitative argument for the dual Meissner effect. The celebrated work by Seiberg and Witten \cite{SW} beautifully realized an abelian dual Meissner effect in a perturbed ${\cal N}=2$ supersymmetric gauge theory. Later, by exploiting a Pouliot-type duality \cite{PouliotType, Kawano}, Strassler argued that a non-abelian version of the dual Meissner effect occurs frequently  \cite{Strassler1,Strassler2}. Non-Abelian dual Meissner effect was also shown to occur quite generally \cite{Konishione,Konishitwo} in many of the 
supersymmetric vacua of the softly broken ${\cal N}=2$ gauge theories coupled to fundamental matter hypermultiplets \cite{SWflavor}.  

%  For the sake of logical clarity, we adopt throughout this paper the point of view that confinement %is caused, in any gauge theory, by condensation of some magnetically charged fields,  whether or %not a finite-mass particle associated to such a field is present in the theory. 

It is believed that those (oblique)  confining vacua have mass gap \footnote{In the literature the term oblique confinement is used generally when the condensing field has a dyonic charge, rather than to a simple magnetic monopole charge. 't Hooft,  however,  used \cite{tHooftConf} 
 the word ``oblique confinement'' with a different meaning. In most of the paper we shall follow the custom and use it in the former sense, except in Section~\ref{oblique} and Subsection~\ref{obliquebis} where we shall use  't Hooft's original usage.  
 }. Classification of such massive vacua in gauge theories were initially studied by 't Hooft \cite{tHooft}. A nice explanation for the classification was given by Donagi and Witten \cite{DW}, which we will follow below. Suppose that the gauge group is $SU(N)$ and the matter contents are trivial under the center $\mathbb{Z}_N$ of the group. External electric charges that one can use to probe the theory are classified by the charge of the center because massless gluons screen most of the charges in various representations of the $SU(N)$ group. On the other hand, the magnetic charges are labelled by the fundamental group: Since the theory includes only adjoint fields, the global structure of the electric group is $SU(N)/\mathbb{Z}_N$. Thus the fundamental group is nonzero $\pi_1(SU(N)/\mathbb{Z}_N)=\mathbb{Z}_N$ and the  monopoles are labelled by $\mathbb{Z}_N$.   The Wilson-'t Hooft loops corresponding to these charges are also classified by $\mathbb{Z}_N\times \mathbb{Z}_N$. One of the striking facts shown by 't Hooft \cite{tHooft} is that finding an order $N$ subgroup in $\mathbb{Z}_N\times \mathbb{Z}_N$ is choosing a corresponding massive vacuum of the theory (including Higgs vacua). The charges in the subgroup specifies the condensed charges in the vacuum. As for (oblique) confinement phase, there are only $N$ types of vacua in which charges corresponding to $W^k H$, $k=0,\cdots N-1$ are condensed respectively. It is well known that ${\cal N}=1$ $SU(N)$ supersymmetric pure gauge theory has $N$  vacua and adjacent vacuum are connected by a  $2 \pi$ shift of the theta angle. Since in a monopole background, the shift of theta angle generates an electric charge \cite{Witteneffect}, it is plausible to think that the  $N$-types of  massive vacua are precisely realized once in each of the $N$ vacua of the theory \cite{CSW}. For example in the $r$-th vacuum, the charges corresponding to $W^r H, (W^r H)^{2}, \ldots, $ are all condensed. The corresponding  Wilson-'t Hooft loop exhibits a perimeter law.

In generalizing the argument to a wide class of supersymmetric gauge theories with different gauge groups, the story will be somewhat modified. As one can see in Table \ref{tab}, the dual coxeter numbers (number of vacua) for classical and exceptional groups are in general different from the center of the groups. Thus, some of massive vacua have to show the same behaviors with respect to the Wilson-'t Hooft loops. To understand the behaviors in various gauge theories we have to study the Witten effect for the Wilson-'t Hooft loops more carefully.  In Appendix, making use of the weight vectors, we shall study the Witten effect for various vacua in pure ${\cal N}=1$ supersymmetric gauge theories. For gauge theories with simply-raced groups and $USp(2N)$ group with $N=$odd, both oblique confinement and confinement phase occur. On the other hand, $Spin(odd)$ and $USp(2N)$ with $N=$even gauge theories have only confinement  phase.   

%%%%%%%%%%%%%%%%%%%%%%%%%%%%%%%%%%%%%%%%%
\begin{table}[htbp]
\begin{center}
\begin{tabular}
{c|c|c|c|c}
\hline
\hline
 group & $G$ & $G^{\vee}$ &  $C(G)$   & dual coxeter {\small \#}  \\
\hline
\hline
$A_{N-1}$ & $SU(N)$ & $SU(N)/\mathbb{Z}_N$ & ${\mathbb Z}_N$ & $N$ \\
    & $SU(MN)/\mathbb{Z}_{N} $ & $SU(MN)/\mathbb{Z}_M$  & ${\mathbb Z}_M$  & $M N$ \\
$B_{N}$ & $SO(2N+1)$ & $USp(2N)$ & ${\bf 1}$ & $2N-1$ \\
$C_{N}$ & $USp(2N)$ & $SO(2N+1)$ & ${\mathbb Z}_2$ & $N+1$ \\
$D_{N}$ & $SO(2N)$ & $SO(2N)$ & ${\mathbb Z}_{2}$ & $2N-2$ \\
$E_{6}$ & $E_6$ & $E_6/\mathbb{Z}_3$ & ${\mathbb Z}_3$ & $12$ \\
$E_{7}$ & $E_7$ & $E_7/\mathbb{Z}_2$ & ${\mathbb Z}_2$ & $18$ \\
$E_8$ & $E_8$ & $E_8$ & ${\bf 1}$ & $30$ \\
$F_{4}$ & $F_4$  & $F_4$ & ${\bf 1}$ & $9$ \\
$G_{2}$ & $G_2$ & $G_2$ & ${\bf 1}$  & $4$ \\ 
\hline
\hline
 $Spin(2N+1)$& & $USp(2N)/\mathbb{Z}_2$ & ${\mathbb Z}_2$ & $2N-1$ \\
 $Spin(2N)$ &$N{\rm :odd}\ge 2$  & $SO(2N)/\mathbb{Z}_2$ & ${\mathbb Z}_4$ & $2N-2$  \\ 
$Spin(2N)$ &$N{\rm :even}\ge 2$&   $SO(2N)/\mathbb{Z}_2$ & ${\mathbb Z}_2\times {\mathbb Z}_2$ & $2N-2$ \\
\hline
\hline
\end{tabular}
\end{center}
\caption{\small GNOW dual group, $C(G)$ (center of $G$)  and the dual coxeter number of the classical and exceptional groups. In all cases,
$\pi_{1}(G^{\vee}) =  C(G)$, and vice versa. }
\label{tab}
\end{table}
%%%%%%%%%%%%%%%%%%%%%%%%%%%%%%%%

In more general theories containing Higgs fields in various representations of the gauge group, we need an extra care for electric screening caused by the Higgsing and  by  magnetic screening induced by the spontaneous nucleation of nonabelian monopoles.
%Such extra screenings considerably reduce possible types of external charges for probing the %vacuum. For example, in an $SU(N)$ gauge theory with an adjoint Higgs $\Phi$  \cite{CSW}, the %standard classification is given by $\mathbb{Z}_N\times \mathbb{Z}_N$.   However,   in more %general types of theories (such as those with a nontrivial superpostential ${\cal W}(\Phi))$, %depending on the dynamics and on the 
 %symmetry breaking pattern, one may end up  having  fewer  available external charges, e.g.,   $%\mathbb{Z}_{\, t}\times \mathbb{Z}_{\, t}$. In such a case,  the Wilson loop   $W^{i}$ will  show  %area law only up to $i=t-1$, rather than up to $N-1$.   
 %This $t$ is {\it the confinement index} introduced by Cachazo, Seiberg and Witten \cite{CSW}, %the main quantity of interest in this paper. 
More concretely let us introduce a mass scale $\Delta$  much larger than the dynamical scale $\Lambda$ of the theory. Suppose that at the scale $\Delta$ a Higgs field gets a vacuum expectation value (vev) and the original gauge group $G$ is spontaneously broken as,
\begin{eqnarray}
G\to G_1 \times G_2 \times \cdots \times G_k.
\end{eqnarray}
The breaking pattern depends on the representation of the Higgs field, on the superpotential, and on the vacuum chosen. A well-studied case has an Higgs field in the adjoint representation of an $SU(N)$ gauge group. Such models were geometrically engineered \cite{CIV,CV} and studied in various aspects such as a gravity dual description \cite{CIV,CV}, a correspondence with a matrix model \cite{DV} and with  generalized Konishi anomaly relations \cite{CDSW}.   Under the symmetry breaking, the matter fields get masses of order $\Delta$. Below the scale, all the massive modes are integrated out. This classical argument is reliable since we assume $\Delta \gg \Lambda$. Finally we are left with a pure ${\cal N}=1$ supersymmetric gauge theory with  gauge group $\prod G_i$, each of which is supposed to be confining. A vacuum of the full $G$ theory is specified by choosing one vacuum for each sub-sector:  we shall label it as $(r_1,r_2,\cdots, r_k)$ where $r_i$ runs from $1$ to the dual coxeter number of $G_i$. Although in each vacuum of sub-sectors, $G_i$ theories are confining, it does not necessarily mean that the full theory is also confining. For instance, as shown in \cite{CSW}, some vacua such as the one with the breaking $SU(N)\to SU(N_1)\times SU(N_2)\times U(1)$ have unconfined $U(1)$ charges  and are in a Coulomb phase. It would be useful to introduce a parameter which specifies if the underlying gauge theory is confining or not. Consider some power of the Wilson loop of the fundamental (or spinor) representation of the original group, which  we denote by  $(W_{\bf fnd})^k$. If these Wilson loops show area law only for some $k$,   $k>1$, then such vacuum is in a confinement phase.  {\it Confinement index is defined to be the smallest positive integer   $k$ for which  $(W_{\bf fnd})^k$ does not show area law} \cite{CSW}.   If  $G_i$ contains a $U(1)$ factor, then $t=1$ and the theory is in a Coulomb phase. On the other hand, if a $U(1)$ is not contained as a factor in the low-energy gauge group, then $t=1$ does not mean the Coulomb law. It may be a perimeter law or a free electric law. Therefore in this case, one cannot conclude weather or not confinement is occurring in the vacuum by the above arguments only. 

In computing the confinement index we must account for all possible types of screening and all conditions for the Wilson-'t Hooft loops. If neither  $W^p$ nor  $W^q$ show area law,    the confinement index is at most  the  greatest common divisor of $(p,q)$.  Thus the index must be a divisor of the dimension of the center of the full gauge group. For the gauge group with a trivial center such as $E_8$, $F_4$ and $G_2$, the confinement index is necessarily unity and there are no external charges to probe a vacuum.

With these knowledges, we can rephrase the classification of vacua of the full theory: On the phase with confinement index $t$, magnetic or dyonic charges corresponding to an order $t$ subgroup $\mathbb{Z}_{\, t}$ in $\mathbb{Z}_N\times \mathbb{Z}_N$ are condensed. When the order $t$ is smaller than $N$,  massless gauge fields can exist and the vacuum does not necessarily have mass gap \cite{DW}. The behavior of the Wilson-'t Hooft loops for the full theory in the vacuum can be determined by the branching rules and the behavior of Wilson-'t Hooft loops for the low-energy gauge groups $\prod G_i$. 

The notion of the confinement index does not necessarily depend on supersymmetry. Nevertheless, we shall focus our attention to supersymmetric gauge theories in this paper for the following reasons: As we will argue in Appendix, we can use various knowledges to understand the behavior of the Wilson-'t Hooft loops in each vacuum of pure ${\cal N}=1$ supersymmetric theories. For non-supersymmetric case this task would be much harder. Another reason is that in supersymmetric vacua the index is protected and is not affected by the change of the parameters. Therefore, in principle one can reproduce the same number from a dual description in string theories or a better low-energy description such as the Seiberg-Witten theory \cite{SW}. We can check our results by making use of such known techniques. As has been well argued, supersymmetric vacua in ${\cal N}=1$ gauge theories can be described by the behavior of flux on a Calabi-Yau manifold. Thus, the index we will compute in this paper may have potential applications for labeling the landscape of field theories realized in string theories. 

Section \ref{general} is devoted to a general consideration of  electric and magnetic screenings under  Higgsing of the gauge group. In Section 3, we compute the confinement index in various systems with different gauge groups.  Section 4 is devoted to the discussion of the condensed charges. We recall briefly 't Hooft's
oblique confinement in the standard (non supersymmetric) $SU(N)$ Yang Mills theory in Section 5.  Section 6 is a brief summary of the lessons from softly broken ${\cal N}=2 $  supersymmetric theories where many of the phenomena discussed in the text are explicitly realized.  We conclude (Section~\ref{concl}) with a brief summary and a further discussion. 
In Appendix we discuss the Witten effect on the  Wilson-'t Hooft loops in supersymmetry preserving vacua for various ${\cal N}=1$ pure gauge theories.

%%%%%%%%%%%%%%%%%%%%%%%%%%%%%%%%%%
\section{General discussion \label{general}} 
%%%%%%%%%%%%%%%%%%%%%%%%%%%%%%%%%%

%%%%%%%%%%%%%%%%%%%%%%%%%%%
\subsection{Electric screening}
%%%%%%%%%%%%%%%%%%%%%%%%%%%

Consider a supersymmetric gauge theory with gauge group $G$. The matter fields are all in the  adjoint representation. To diagnose the theory, introduce a Wilson loop for a representation ${\cal R}$ in $G$. Gluons in the theory can combine with the external charge in the representation ${\cal R}$ and neutralize the charge. This is the so-called electric screening.  
Electric screening turns one representation into another. Since the adjoint representation is trivial under the center, the tensor product of a representation ${\cal R}$ with adjoint fields yields various representation with the same $N$-ality,
$${
{\cal R}\otimes (adj)=\sum {\cal R}_i.
}$$
As fields in the adjoint representation do not carry charges for the center of the group,  however,  the latter charges are not screened. Therefore the Wilson loop for various representations can be labelled by the charge of the center (we call it the ``$N$''-ality). It is thus useful to consider a Wilson loop which is the $r$-th tensor product of the one in the fundamental representation,   rather than considering the Wilson loop for each  irreducible representation  separately.     The Wilson loop for such a direct-product representation can be written simply as 
$${
W_{{\cal R}\otimes \cdots \otimes {\cal R}}=(W_{\cal R})^r.
}$$   

Now consider introducing a Higgs field in a generic irreducible representation of $G$ in the theory. If this field carries a nontrivial charge under the center, then available external charges get  reduced because of extra screening. Suppose that such a Higgs field gets vev and symmetry breaking $G\to \prod G_i$ takes place  at the energy scale $\Delta$ which is much bigger than the dynamical scale $\Lambda$ of the underlying  $G$ theory. The breaking pattern depends on the representation of the Higgs field. After integrating out the massive matters, we are left with a pure ${\cal N}=1$ supersymmetric gauge theory with gauge group $\prod G_i$ \footnote{In most of the following we assume that the gauge symmetry breaking is not  accompanied by a global symmetry breaking, which would imply the presence of massless Nambu-Goldstone bosons.}.   At low energies, nonabelian parts in $G_i$ gauge theories are supposed to be confining. In each vacuum of  $G_i$ theory the Wilson loop exhibits an area law. However,  even if the sub-sectors  are confining, the full theory is not necessarily confining. To diagnose that, we use the Wilson loops of the full $G$ theory. Suppose that  the fundamental representation of the group $G$ is decomposed into ${\bf fnd\to \sum A_j}$ where ${\bf A_j}$ are an irreducible representation of $\prod G_i$. If the $r$-th tensor product of the fundamental representation includes a singlet of the low-energy group $\prod G_i$, then the Wilson loop will not exhibit area law:
% because the charges for the broken group are not confined and the singlet does not feel confinement in sub-theories. Therefore we have
$${
(W_{\bf fnd})^r= \bigg(\sum_j  W_{\bf A_j} \bigg)^{r}=  W_{\bf singlet}+\cdots \simeq \ {\rm no\ area\ law},
}$$
where $W_{\bf A_j}$ are Wilson loops for sub-theories. If there is a term which does not show area law in the right hand side, the full Wilson loop $(W_{\bf fnd})^r$ also does not show area law because a term with no-area law dominates over  the ones with area law. 

%%%%%%%%%%%%%%%%%%%%%%%%%%%%%%%%%%%
\subsection{Magnetic screening}
%%%%%%%%%%%%%%%%%%%%%%%%%%%%%%%%%%

Under the symmetry breaking $G \to H$, with a  non-trivial $\pi_2(G/H)$, 
 soliton magnetic monopoles are generated  and become  part of the spectrum of the low-energy theory.  These monopoles belong to various  representations of the GNO dual group \footnote{We shall not discuss here the subtleties around the notion of ``non-Abelian monopoles'' and the related difficulties in defining the continuous GNO duality transformations quantum mechanically.   For the purpose of the present discussion, 
their existence, not their detailed properties, matters.
} of $H$ (which we denote by  $H^{\vee}$). Although such monopoles are in general massive, they can screen a sufficiently large 't Hooft loop \cite{CSW}.  Suppose, for simplicity, that the unbroken group is $G_1\times G_2$ and a magnetic monopole screens $H_1^m H_2^n$ where $H_1$ and $H_2$ are 't Hooft loops corresponding to the fundamental (or spinor) representations of dual group $G_1^{\vee}$ and $G_2^{\vee}$. In the simple case of $SU(N)$  theory broken to $SU(N_{1})\times SU(N_{2})\times U(1)$  
illustrated in  \cite{CSW},  for instance, the minimum monopoles have the quantum numbers of $H_{1}\, H_{2}^{-1}$, as each monopole resides in one of the  broken $SU(2)$
subgroups  embedded in $SU(N)/SU(N_{1})\times SU(N_{2})$. 
As in the previous section, consider the decomposition of a Wilson loop of the full $G$ theory, 
$${
(W_{\bf fnd} )^r\simeq \sum_{i} W^{a_i}_1 W_2^{b_i}.
}$$
where $\simeq$ means up to $N$-ality. 
Spontaneous nucleation of the massive monopoles from the vacuum  may screen the Wilson loop,
\begin{eqnarray}
\label{bun} (W_{\bf fnd} )^r\simeq \sum_{i} (W^{a_i}_1 H_1^m )(W_2^{b_i}H_2^n).
\end{eqnarray}
By definition in each vacuum of $G_i$ theories, some of the Wilson-'t Hooft loops show no area law. Suppose that $W^{r_1}_1H_1$ and $W^{r_2}_2H_2$ show no area law in a vacuum. In this case if \eqref{bun} includes a term with a pair $\{(a_i,m)=(k r_1, k),  (b_i,n)=(l r_2, l)  \}$ where $k,l$ are integers, then $(W_{\bf fnd} )^r$ will not show area law. This is the magnetic screening caused by magnetic monopoles. 

To understand the effects of  magnetic screening we need to know the representation of nonabelian monopoles. This nontrivial task has been studied in literatures. In \cite{WeinbergYi,KonishiMurayama} the minimal magnetic monopoles arising in spontaneously broken classical and exceptional gauge groups with an adjoint Higgs field were constructed. Irreducible representations of the minimal monopoles in the dual groups were explicitly shown there, which will be needed in section three. Interestingly, all such representations can be understood in terms of the Montonen-Olive duality of ${\cal N}=4$ theories \cite{MOdual}. Because of the self-duality, the representation of the monopoles under the dual group must be the same as that of the massive vector bosons with respect to the original electric group.  This is not so surprising because the monopole solution constructed in \cite{WeinbergYi,KonishiMurayama} can be embedded into ${\cal N}=4$ theories.

However, in models with Higgs fields in non-adjoint representations, the Olive-Montonen duality does not hold. Non-abelian monopoles for various non-adjoint Higgs models were constructed in \cite{Bais}. As is well known, the problem of finding exact monopoles solutions is greatly simplified by use of the Bogomolny equations if the Higgs fields transform under the adjoint representation. However in non-adjoint Higgs models, such a construction breaks down. Nevertheless, Bais and Laterveer managed to find the monopole solutions for various models by suitably choosing the subgroups and by solving the Bogomolny equations for the latter.

%%%%%%%%%%%%%%%%%%%%%%%%%%%%%%%%%%%%%%%%%%%
\subsection{The confinement index and $\theta$ periodicity}
%%%%%%%%%%%%%%%%%%%%%%%%%%%%%%%%%%%%%%%%%%%

Accounting for the electric and magnetic screenings we can obtain Wilson loops for the full theory which do not show an area law, which we denote by $W^{k_i}$, $i=1,2,\cdots$. As argued in Introduction, the smallest number $t$ such that $W^t$ does not show an area law is given by the greatest common divisor $t=GCD(k_1,k_2,\cdots)$, which is the confinement index introduced in \cite{CSW}. In general, this index depends on the rank of the low-energy gauge group, $rank(G_i)$, and the labels of the vacua $r_i$. The $r_i$ dependence comes from the magnetic screening by a nonabelian monopole generated by the Higgsing. As a nontrivial consistency check of the dependence, one can use invariance of the confinement index under the $2\pi$ shift of the theta angle of the original $G$ gauge theory. When we vary the theta angle, the ones for the low-energy $G_i$ theories also get shifted accordingly, which depends on the embedding of the small group $G_i$. It is convenient to use the index of embedding to understand the behavior \cite{Bais,MurayamaSaki,Slansky}. Suppose a small group $H$ is embedded in $G$ and a representation ${\cal R}$ of the group $G$ has a decomposition under the $H$ subgroup ${\cal R} \to \sum_i^k{\cal R}_i$. The $\mu_{\cal R}$ is defined by the generators $T_a$ of the representation ${\cal R}$ as ${\rm Tr}_{\cal R}\, T_aT_b =\mu_{\cal R} \delta_{ab}$. We normalize the generators such that the Dynkin indices of the fundamental representations of $SU(N)$ and $USp(2N)$ are one and those of the vector representations of $SO(N)$ are two. As for exceptional groups we normalize such that $\mu_{\bf 27}=6$ for $E_6$, $\mu_{\bf 56}=12$ for $E_7$ and $\mu_{\bf 7}=2$ for $G_2$. Now let us define the index of embedding $J$,
\begin{eqnarray}
J =  \frac{\sum_i\mu_{{\cal R}_i}}  {\mu_{\cal R}}. \label{indexofembedding}
\end{eqnarray}
This index is an integer and independent of the choice of the representation ${\cal R}$. If the index $J$ is bigger than one, then the matching relation of the gauge couplings of the high and low-energy theories is given by
\begin{eqnarray}
\left(\frac{\Lambda^{b_G}}{v^{b_G}} \right)^{J}=
\frac{\Lambda^{b_H}}{v^{b_H}}.
\end{eqnarray}
Thus, under $2\pi$ shift of the theta angle of the $G$ theory, the theta angle for the small group $H$ shifts $2\pi J$. Once we understand the appropriate behavior of the theta angle, we can check if the confinement index is invariant under the shift, which provides  us with a rather nontrivial check of the magnetic screening mechanism.

%%%%%%%%%%%%%%%%%%%%%%%%%%%%%%%%%%
\section{Computation of Confinement Index \label{confind}}
%%%%%%%%%%%%%%%%%%%%%%%%%%%%%%%%%%

%%%%%%%%%%%%%%%%%%%%%%%%%%%
\subsection{$USp(2N)$ with antisymmetric tensor Higgs \label{USpAnti}}
%%%%%%%%%%%%%%%%%%%%%%%%%%%

As a first example we consider an ${\cal N}=1$ supersymmetric $USp(2N)$ gauge theory with an antisymmetric tensor Higgs field. By turning on a tree-level superpotential for the Higgs, one can break the gauge group  as  
$${
USp(2N)\to  \prod_i USp(2N_i) \qquad \sum_i N_i=N.
}$$
Since $\pi_2(G/H)$ of this symmetry breaking is trivial, there are no 't Hooft-Polyakov monopoles in this system. The fundamental representation of the $USp(2N)$ group is decomposed as \cite{Slansky}
$${
{\bf 2N\to (2N_1,1,\cdots ,1)+ (1, 2N_2,1\cdots ,1)+\cdots +(1,\cdots, 2N_k ,1,\cdots)+\cdots},
}$$
thus the Wilson loop for the fundamental representation breaks into the sum of the ones for each $USp(2N_i)$ groups, $W_{\bf 2N}\to \sum_{i=1}^N W_{\bf 2N_i}$. Invariant tensor in symplectic group tells us that $(W_{\bf 2N})^2$ does not show area law. Accounting for the arguments in the previous section we conclude that in this case, there are no extra electric/magnetic screenings caused by the Higgsing. The Wilson loop for the fundamental representation $W_{\rm 2N}$ shows area law. Thus the confinement index in this system is $t=2$. No matter how we choose vacua in each $USp(2N_i)$ theory, all vacua for $USp(2N)$ theory are confining or oblique confinement phase. This is consistent with the argument in the work  \cite{Freddy} where Cachazo made a map of the generalized Konishi anomaly equations \cite{CDSW,CSW} for $U(N)$ theory with adjoint Higgs and the ones in  a $USp(2N)$ theory with an antisymmetric tensor. The vacua in $USp$ theory maps to the one in $U(N)$ theory with confinement index two. Our direct calculation of the confinement index supports the result.

%%%%%%%%%%%%%%%%%%%%%%%%%%%%%%%%%%%%%%%%%%%
\subsection{$USp(2N)$ with adjoint Higgs}
%%%%%%%%%%%%%%%%%%%%%%%%%%%%%%%%%%%%%%%%%%%

As the next example, consider supersymmetric $USp(2N)$ gauge theory with an adjoint Higgs. For simplicity, we focus on the following three breaking patterns. These breaking patterns have been well studied in various contexts such as geometric transition \cite{SO} and factorizations of Seiberg-Witten curves \cite{AFO1}. 

\vspace{0.2cm}
%%%%%%%%%%%%%%%%%%%%%%%
$\bullet$ {$USp(2N)\to (SU(N)\times U(1))/\mathbb{Z}_N$}
%%%%%%%%%%%%%%%%%%%%%%%

\noindent
An external charge which is not screened by the adjoint fields is the center $\mathbb{Z}_2$. Our main concern is the behavior of Wilson loop for the fundamental representation of $USp(2N)$ group. Since the branching rule of the representation is ${\bf 2N \to N\oplus \bar{N}}$ \cite{Slansky}, the Wilson loop decomposes into $W_{\bf 2N}\to W_{1}+ W_{1}^{-1}$, where the $W_1$ is in the fundamental representation of the $SU(N)$ group. As argued in \cite{KonishiMurayama} under this symmetry breaking, a non-abelian monopole in the fundamental representation of the dual gauge group $H^{\vee}=U(N)$ is generated. Spontaneous nucleation from the vacuum  of the massive monopole pair  provides screening of the Wilson loop $W_{\bf 2N}$. To see that, pick the $r$-th vacuum of the low-energy  $SU(N)$ gauge theory and consider the Wilson loop for the $r$-th tensor product, 
\begin{eqnarray}
(W_{\bf 2N})^{r}\simeq (W_{1})^{r}+\cdots =(W_{1})^{r}H_{1}+\cdots .
\end{eqnarray}
In the last equality we used the fact that the generated monopole screens 't Hooft loop $H_1$ corresponding to the fundamental representation in dual gauge group $H^{\vee}$. 
Since in the $r$-th vacuum $W_1^r H_{1}$ does not show area law, neither does the full Wilson loop $(W_{\bf 2N})^r$. Electric screening caused by the Higgsing gives a condition for the Wilson loop. To see another condition, take the Wilson loop for the $N$-th tensor product of the fundamental representation. This representation includes a singlet because of the invariant tensor epsilon in $SU(N)$ group. Therefore the Wilson loop for the representation does not show area law. The confinement index for this symmetry breaking is summarized as $t=GCD(2,N,r)$, i.e.,  either $t=1$ or $t=2$.   When the index is one, the Wilson loop $W_{\bf 2N}$ does not show area law.  However in the IR, there is an unbroken $U(1)$ group which is a subgroup of the original gauge group. The $U(1)$ charge exhibits  Coulomb law. Thus the vacua with $t=1$ is in a Coulomb phase.

\vspace{0.2cm}
%%%%%%%%%%%%%%%%%%%%%%%
$\bullet$ {$USp(2N)\to USp(2N-2)\times U(1)$}
%%%%%%%%%%%%%%%%%%%%%%%

\noindent
In this case, the decomposition of the fundamental representation contains a singlet of the $USp(2N-2)$ group. The Wilson loop $W_{\bf 2N}$ therefore does not exhibit an area law. The confinement index is $t=1$. 
%As in the previous breaking pattern,
 The Wilson loop always shows Coulomb law because of the unbroken $U(1)$ gauge group.

\vspace{0.2cm}
%%%%%%%%%%%%%%%%%%%%%%%
$\bullet$ {$USp(2N)\to USp(2N-2M)\times U(M)$}
%%%%%%%%%%%%%%%%%%%%%%%

\noindent
Again let us start with the branching rule for the fundamental representation representation of the $USp(2N)$ group,
$${
{\bf 2N \to (2N-2M,1)+ (1,M)+(1,\bar{M})}. 
}$$
Since the tensor product of $M$ copies of the representation contains a singlet, $(W_{\bf 2N})^{M}$ show no area law. As argued in \cite{KonishiMurayama}, a nonabelian monopole in  $({\bf 2N-2M+1,M})$ representation of dual gauge group $H^{\vee}=Spin(2N-2M+1)\times U(M)$ is generated  under this symmetry breaking. This monopole screens the 't Hooft loop $H_1^2H_2$ where $H_1$ is the one for spinor representation of $Spin(2N-2M+1)$ and $H_2$ is the one for the fundamental representation of $U(M)$. Because of magnetic screening we see that $(W_{2N})^{2r_1+r_2}$ does not exhibit area law, when we are in the $r_1$-th vacuum of the pure $USp(2N-2M)$ theory and in the $r_2$-th vacuum of the pure $SU(N)$ theory at the same time. Thus the confinement index is $t=GCD(2,M,2r_1+r_2)$ or equivalently $t=GCD(2,M, r_2)$. The fact that the index depends only on $r_2$ is somewhat surprising because the index appears to be changed by $2\pi $ shift of the theta angle of the original $USp(2N)$ theory. However, looking at the matching relation \cite{IntriligatorShigemori} carefully, we see that the $U(M)$ has the index of embedding two,
$${
\Lambda_{USp}^{3(2N-2M+1)}=C_{USp}\cdot \Lambda^{2(2N+1)}, \qquad \Lambda_{U}^{3M}=C_{U}\cdot \Lambda^{4(2N+1)}.
}$$
where $\Lambda$ is the dynamical scale of the original gauge theory $USp(2N)$ and $C_{USp}$ and $C_{U}$ are constants depending on the parameters in superpotential. Therefore under the $2\pi$ shift of the theta angle, in $U(M)$ theory, its theta angle is shifted by $4\pi$ namely, $r_2\to r_2+2$. Thus the index is invariant. Moreover this index is also consistent with the factorization solutions of Seiberg-Witten curves \cite{AFO1}. A $USp(4)$ theory with a quartic superpotential of an adjoint field has one branch in which two kinds of vacua with breaking patterns $USp(2)\times U(1)$ and $U(2)$ exist. Such vacua must have the same confinement index. Our calculation shows that the index for the former one is $t=1$. For the latter one, one of the two vacua that is in the same branch has indeed  $t=1$. As another example, a branch for $USp(6)$ theory contains vacua with three breaking patterns $USp(2)\times U(2)$, $USp(4)\times U(1)$ and $U(3)$. All vacua in this branch have the same confinement index,  $t=1$. 

%%%%%%%%%%%%%%%%%%%%%%%%%%%%%%%%%%%%%%%%%%%
\subsection{$E_6$ with adjoint Higgs}
%%%%%%%%%%%%%%%%%%%%%%%%%%%%%%%%%%%%%%%%%%%

Consider now the confinement index for an $E_6$ gauge theory with an adjoint Higgs. The low-energy dynamics of this theory such as the glueball superpotential and the generalized Konishi anomaly has not been explored. It would be interesting to reproduce our results by using such techniques. As an illustration, we focus on two breaking patterns.

\vspace{0.2cm}
%%%%%%%%%%%%%%%%%%%%%
$\bullet$ {$E_6 \to  (SU(6)\times U(1))/\mathbb{Z}_2$ \label{E6II}}
%%%%%%%%%%%%%%%%%%%%%

\noindent
The branching rule for the fundamental representation of $E_6$ group is ${\bf 27\to \bar{6}+\bar{6}+15}$ \cite{Slansky}. Using the invariant tensor in $SU(6)$ group we know that $W_{\bf 27}^6$ does not show area law. This does not yield new constraint because the center of $E_6$ is $\mathbb{Z}_3$ and we know already  that $W_{\bf 27}^3$ does not show area law. According to \cite{WeinbergYi}, non-abelian monopole generated by this breaking  is in a rank-three antisymmetric tensor representation of the dual gauge group $H^{\vee}=U(6)$. Therefore the monopole screens the  't Hooft loop $H_1^3$ where $H_1$ is the 't Hooft loop corresponding to fundamental representation of the dual group. However magnetic screening for Wilson loop $W_{\bf 27}$ does not give us a new condition. We conclude that the confinement index is $t=3$ and all vacua of $E_6$ with this breaking pattern are confining or oblique confinement phase.

\vspace{0.2cm}
%%%%%%%%%%%%%%%%%%%%%%%
$\bullet$ {$E_6\to ( Spin(10)\times U(1) )/\mathbb{Z}_4$}
%%%%%%%%%%%%%%%%%%%%%%%

\noindent
The branching rule for  the fundamental representation of $E_6$ includes a singlet of the low-energy gauge group. Therefore, we immediately conclude that all the external charges are completely screened and the confinement index is $t=1$.

%%%%%%%%%%%%%%%%%%%%%%%%%%%%%%%%%%%%%%%%%
\subsection{$SO(odd)$ with adjoint Higgs}
%%%%%%%%%%%%%%%%%%%%%%%%%%%%%%%%%%%%%%%%%

An available charge for probing the theory is the spinor charge. Below we study the behavior of the Wilson loop for this representation, 
in three  relatively simple breaking patterns.

\vspace{0.2cm}
%%%%%%%%%%%%%%%%%%%%%%%%%%%%%%%%%
$\bullet$ {$SO(2N+1)\to SO(2N-1)\times U(1) $}
%%%%%%%%%%%%%%%%%%%%%%%%%%%%%%%%%

\noindent
Since the branching rule for the spinor representation of $SO(2N+1)$ is \cite{Slansky}
$${
{\bf 2^{N}\to 2^{N-1}+2^{N-1}},
}$$
the Wilson loop decomposes into $W_{\bf 2^{N}}\to W_{1}+W_{1}$ where $W_{1}$ is the Wilson loop for the spinor representation of unbroken group $SO(2N-1)$. Note that we omitted the  $U(1)$ charges. Since the product of two of the spinor representation contains a singlet of $SO(2N+1)$, a Wilson loop $(W_{\bf 2^N} )^2$ shows no area law. Since the product of two of the spinor representation ${\bf 2^N}$ contains a singlet of $SO(2N+1)$, a Wilson loop $(W_{\bf 2^N})^2$ shows no area law. Since, by our assumption, the pure $SO(2N-1)$ gauge theory is confining, Wilson loop $W_1$ shows area law, and the index is $t=2$. 

Next consider magnetic screening caused by nonabelian monopoles generated by this symmetry breaking. From \cite{KonishiMurayama}, we see that there is a nonabelian monopole in the fundamental representation of dual group $H^{\vee}=USp(2N-2)\times U(1)$. Therefore by spontaneous nucleation of the monopole, 't Hooft loop $H_1$ in the fundamental representation of the dual group $H^{\vee} $ is screened, and we can see that 
$${
W_{\bf 2^{N}} \to W_{1} +W_{1}= W_1 H_1 + W_1H_1.
}$$
Since the Wilson loop $W_1$ shows an area law and the 't Hooft loop $H_1$ does not show an area law, $W_1 H_1$ must show an area law. Therefore, taking into account Witten effect in Appendix, we conclude that all vacua in the pure $SO(odd)$ theory must be in a confinement phase,  not in an oblique confinement phase. It gives strong support for our claim in Appendix, and is also consistent with Witten's argument \cite{WittenIndex}\footnote{We would like to thank T. Kawano for suggesting us this argument.}.

\vspace{0.2cm}
%%%%%%%%%%%%%%%%%%%%%%%%%%%%%%%%%
$\bullet$ { $SO(2N+2M+1)\to SO(2N+1)\times (SU(M)\times U(1))/\mathbb{Z}_M$} 
%%%%%%%%%%%%%%%%%%%%%%%%%%%%%%%%%

\noindent
The branching rule for the spinor representation for the breaking is given by \cite{Slansky}
\begin{eqnarray}
&&{\bf 2^{M+N} \to } \sum_{j=0}^k  \big({\bf 2^{N}} ,\,  [2j+1]\big)+ \sum_{j=0}^k\big({\bf 2^{N}},\, [2j]\big) \,\quad {\rm for}\ \ M={ 2k+1} , \nonumber \\
&&{\bf 2^{M+N} \to } \sum_{j=0}^{k-1}  \big({\bf 2^{N}} ,\,  [2j+1]\big)+ \sum_{j=0}^k \big({\bf 2^{N}},\, [2j]\big) \,\quad {\rm for}\ \ M= 2k, \nonumber
\end{eqnarray}
where $[s]$ is rank $s$ antisymmetric tensor representation of $SU(M)$ group whose dimension is ${\rm dim}[s]=\/ \/{}_MC_{s}$. Therefore the decomposition of  the Wilson loop in spinor representation of $SO(2M+2N+1)$ group is given by
$${
W_{\bf 2^{M+N}} \to \sum_{j=0}^M\, W_1\, W_2^j .
}$$
According to \cite{KonishiMurayama}, a magnetic monopole generated in this breaking belongs to the bifundamental representation of the dual gauge group $H^{\vee}=USp(2N)\times U(M)$. Thus it screens the  't Hooft loop $H_1 H_2$ where $H_1$ and $H_2$ are 't Hooft loops corresponding to the fundamental representations of $USp(2N)$ and $U(M)$,  respectively. Again we use the fact that all vacua in $SO(2N+1)$ theory are in confinement phase and the condensed charge is the one for $H_1$. Thus the spinor charge of the original group remains unscreened and we conclude the index is $t=2$.

\vspace{0.2cm}
%%%%%%%%%%%%%%%%%%%%%%%%%%%%%%%%%
$\bullet$ {$SO(2N+1)\to ( SU(N)\times U(1) )/\mathbb{Z}_{N}$ }
%%%%%%%%%%%%%%%%%%%%%%%%%%%%%%%%%

\noindent
In this system, the branching rule for the spinor representation of $SO(2N+1)$ includes a singlet. Thus by electric screening the Wilson loop for the spinor representation shows no area law. We conclude that the index is $t=1$ and  all vacua are in a Coulomb phase.

This index is consistent with the factorization solutions of Seiberg-Witten curves \cite{AFO1}. A $SO(7)$ theory with a quartic superpotential of an adjoint field has one branch in which two kinds of vacua with breaking patterns $SO(3)\times U(2)$ and $U(3)$ exist. Such vacua must have the same confinement index. Our calculation shows that the index for the latter one is $t=1$. For the former one, two of the four vacua that are in the same branch has indeed  $t=1$. Instead, the other two vacua with $SO(3)\times U(2)$ belong to the same branch as the one with the breaking pattern $SO(5)\times U(1)$. All vacua in this branch have the same confinement index,  $t=2$. These two branches originate from the two different types of the factorization forms of the Seiberg-Witten curve. As argued in \cite{AFO1}, to extract all possible vacua for a $SO(odd)$ theory with a quartic superpotential we have to consider the following factorization forms,
\begin{eqnarray}
y^2=x^2 (H_{2N-4})^2 F_6(x) \quad {\rm and }\quad y^2= (H_{2N-2})^2 F_4(x). \nonumber 
\end{eqnarray}
Our calculation tells us that for the vacua in the former branch the Wilson loop for the spinor representation exhibits an area law while for the latter one it does not.

%%%%%%%%%%%%%%%%%%%%%%%%%%%%
\subsection{$SU(N)$ with symmetric tensor Higgs}
%%%%%%%%%%%%%%%%%%%%%%%%%%%%

A symmetric tensor field has charge two under the center of $SU(N)$. Therefore, due to the electric screening,  the only available charge for probing vacua is the charge of the fundamental representation. Adding a vev to the Higgs fields one can break the $SU(N)$ gauge group as $SU(N) \to Spin(N)/\mathbb{Z}_{2}$. 
Because of the invariant tensor in $SU(N)$ group, one knows that $W_{\bf N}^N$ does not show area law.    For $N={\rm odd}$,  we  immediately conclude that $t=GCD(N,2)=1$   (the unconfined charges are closed under addition \cite{CSW} and  
$ W^{2m}\sim W^{1}$ for some integer $m$). Stated slightly differently,  the gauge group is truly $SU(N)$, as $2$ and $N$ are relatively prime,  hence no Wilson loop can show area law just as in the $SU(N)$  theory with quarks in the fundamental representation.

The branching rule for the fundamental representation of $SU(N)$  into $Spin(N)/\mathbb{Z}_{2}$    is ${\bf N \to N}$. 
 As argued in \cite{Bais}   a nonabelian monopole is generated under this symmetry breaking. When $N=2k$ the monopole belongs to a vector representation of the dual group $H^{\vee}=Spin(2k)$. Here for simplicity we focus on $k={\rm odd}$. Since the monopole is in the vector representation the  't Hooft loop screened by the monopole is $H_1^2$. Magnetic screening for the Wilson loop $(W_{\bf N})^{r}$ is 
$${
(W_{\bf N} )^r \to (W_1)^{2r} =(W_1^r H_1)^2,
}$$
where $W_1$ is the Wilson loop in the spinor representation of $Spin(2N)$ group. Thus, we conclude that $t=GCD(2,N, r_{1})$, where $r_{1}$ labels the vacua of the low-energy $Spin(2k)$ theory. In summary,  half of the vacua in this system have $t=1$, the other half  $t=2$.
 As a consistency check, let us consider $2\pi$ shift of the theta angle of the $SU(N)$ theory. As was shown in \cite{Bais}, the Dynkin index of embedding of the unbroken group is two. Therefore under the shift the theta angle of the small group $Spin(N)$ gets shifted $4\pi$, namely $r_{1} \to 2+r_{1}$ and  the index remains invariant.

%%%%%%%%%%%%%%%%%%%%%%%%%%%%%%%%%%%%%%%%%
\subsection{$Spin(2M+2N+1)$ with symmetric tensor Higgs}
%%%%%%%%%%%%%%%%%%%%%%%%%%%%%%%%%%%%%%%%%

The last example is $Spin(2M+2N+1)$ with a symmetric tensor Higgs. For simplicity, we focus on the following breaking pattern with $M=2k+1$,
$${
Spin(2M+2N+1) \to \big(Spin(2M)\times Spin(2N+1)  \big)/\mathbb{Z}_2.
}$$
The branching rule for the  spinor is 
$${
{\bf 2^{M+N}\to (2^{M-1}_s, 2^N)+(2^{M-1}_c,2^{N})}.
}$$
Thus the decomposition of the Wilson loop  is 
$${
W_{\bf 2^{M+N}}\to W_{1} W_2 +W_1^{\prime} W_2. 
}$$
where the Wilson loops $W_1$ and $W^{\prime}_1$ are in the representation ${\bf 2_s^{N-1}}$ and ${\bf 2_c^{M-1}}$, respectively of $Spin(2M)$, and $W_2$ is in the representation ${\bf 2^N}$ of $Spin(2N+1)$. Repeating the same discussion as in subsection $3.4$, one can see that the Wilson loop $W_2$ is in the pure $SO(2N+1)$ shows an area law, and the Wilson loop $(W_{\bf 2^{M+N}})$ shows no area law. Thus we find that $t=2$.

According to \cite{Bais}, a nonabelian monopole generated in this breaking pattern is in a  $({\bf 2M, 2N})$ representation of the dual group $H^{\vee} = Spin(2M)\times USp(2N)$. Thus $H_1^2 H_2$ is screened by this monopole where $H_1$ is  the 't Hooft loop corresponding to  the spinor representation of $Spin(4k+2)$ group and $H_2$ is the one for the fundamental representation of $USp(2N)$. However this monopole does not yield new constraint. So we conclude that the confinement index for this breaking pattern is always $t=2$. It would be interesting to check this results by making a map of the generalized Konishi anomaly equations to the ones for $U(N)$ with adjoint theory, following the argument shown in \cite{Freddy}.

%%%%%%%%%%%%%%%%%%%%%%%%%%%%%%%%%%
\section{Condensed charges \label{condch}}
%%%%%%%%%%%%%%%%%%%%%%%%%%%%%%%%%%

So far we have concentrated on the problem of determining the confinement index. 
In this section we study the behavior of various  Wilson-'t Hooft loops for a system with confinement index $t>1$ (i.e., a confining system) more carefully, and attempt to find out the charges which are condensed in each case.

 As for the vacua with $t=1$, none of magnetic charges or dyonic charges are condensed and as a consequence,  no Wilson loops show area law. To understand the behavior of the large Wilson-'t Hooft loops we first need to know the decomposition of 't Hooft loops under the Higgsing. A 't Hooft loop corresponds to a weight vector of a representation of the GNO dual group, so its decomposition can be read off from the branching rule for the representation under the symmetry breaking of the dual gauge group $G^{\vee}\to \prod G^{\vee}_i$. Using the branching rule and the known behavior of Wilson-'t Hooft loops for the pure Yang-Mills theories, one can determine which loops show area law and which charges are condensed. Note that when a Wilson-'t Hooft loop does not exhibit area law, it does not necessarily imply that the corresponding charge is condensed. It may be Coulomb law or free magnetic law. Below, we will illustrate our argument in three examples. It is straightforward to apply it to other systems studied  in the previous section.

%%%%%%%%%%%%%%%%%%%%%%%%%
\subsection{$SU(N)\to SU(N_1)\times SU(N_2)\times U(1)$}
%%%%%%%%%%%%%%%%%%%%%%%%%

As a first example we study $SU(N)$ supersymmetric gauge theory with the breaking pattern $SU(N)\to SU(N_1)\times SU(N_2)\times U(1)$. In a vacuum with confinement index $t$,  $W^{t}$ does not show area law by definition. From this we know that neither  the charges in the fundamental representation of dual group corresponding to the  't Hooft loop  $H$ nor   the dyonic charge $WH$  are condensed. For if $H$ or $WH$ were condensed  $W^t$ would have to show area law, as $H$ and $W^t$ or $WH$ and $W^t$ are relatively non-local.   
Repeating the argument for various charges we can deduce
 that the possible condensed charges in a vacuum with confinement index $t<N$ are the ones associated with $W^A H^{N\over t}$ where $A$ is not determined by this argument only. However by the decomposition  of the large Wilson-'t Hooft loops and by using the fact that in the vacuum some of $W_i^{r_i} H_i$ do not show area law, one can conclude that the possible condensed charges are the ones generated by
$${
W^{kN\over t} H^{N\over t},\qquad k=0,1,\cdots ,t-1.
}$$
Note that $t$ is a divisor of $N$ \cite{CSW}. Clearly such charges form a $\mathbb{Z}_{\, t}$ subgroup in $\mathbb{Z}_N\times \mathbb{Z}_N$. Condensation of the charges causes a dual Meissner effect and $W^l$, $l=1,2,\cdots t-1$ enforced to show area law.

To see more concretely, let us focus on the relatively simple breaking pattern $SU(2n)\to SU(n)\times SU(n)\times U(1)$ and pick vacua with confinement index $t=n$. According to \cite{CSW} there are $n$ such vacua
$${
(r_1,r_2)=(0,0),\, (1,1),\, \cdots ,(n-1,n-1).
}$$
The decomposition of  't Hooft loop for the fundamental representation is given by $H=H_1+H_2$. On the $(k,k)$ vacuum by the decomposition it is easy to see that $W^k H$ does not show area law. However as argued above this does not immediately imply  that such charge is condensed. From the consistency with the fact $W^l$, $l=1,\cdots t-1$ show area law, the condensed charge should be $(W^k H)^{2n\over t}=(W^k H)^n$. Thus this loop shows perimeter law. The others $(W^k H)^l$, $l=1,\cdots n-1$ are expected to show Coulomb law because of the existence of the unbroken $U(1)$ gauge group.

This result is consistent with the Witten effect discussed in Appendix. Although in Appendix we focused on the pure Yang-Mills theories, the argument is applicable to  our present case. To check the consistency, let us consider the adjacent vacuum $(k+1,k+1)$ in which the condensed charge is the one for $(W^{k+1}H)^{2n\over t}$. Since the index of embedding for the small group $SU(n)$ is one, the $2\pi$ shift of the theta angle of $SU(2n)$ theory generates the same amount of shift of the angle for the small $SU(n)$ theories. Therefore we observe the expected behavior of the Wilson-'t Hooft loop under the shift,
$${
(r_1,r_2)=(k,k)\to (k+1,k+1), \qquad (W^kH)^{n} \to (W^{k+1}H)^n.
}$$
From the argument of a weight vector, we knew that under such shift $H$ transforms into $WH$, which is consistent with the above behavior.

%%%%%%%%%%%%%%%%%%%%%%%%%
\subsection{$USp(2N)\to USp(2N_1)\times USp(2N_2)$}
%%%%%%%%%%%%%%%%%%%%%%%%%

The next example is the theory studied in section \ref{USpAnti}. Here for simplicity we consider only the case with $(N_1,N_2)=(odd,odd)$. As argued in Appendix, the number of vacua of $USp(2N_i)$ theory is $N_i+1=even$ and a Wilson-'t Hooft loop transforms as $H_i\to W_iH_i\to H_i$ under the $2\pi$ shifts of the theta angle. There are four types of vacua for the full theory, $(r_1,r_2)=(H_1,H_2)$, $(W_1H_1, H_2)$, $(H_1, W_2H_2)$ and $(W_1H_1,W_2 H_2)$. Let us start with the case $(r_1,r_2)=(H_1,H_2)$. In the breaking of the dual group $Spin(2N+1)\to Spin(2N_1+1)\times Spin(2N_2+1)$, the 't Hooft loop for the spinor representation decomposes into $H=H_1H_2$. Thus we immediately understand that $H$ does not show area law.  As shown in section \ref{USpAnti} all vacua have confinement index two, so the charge for  $H$ has to be condensed and $H$ shows perimeter law. On the other hand, $WH$ shows area law. In the same way, we consider the vacua with $(r_1,r_2)=(W_1H_1,W_2H_2)$. The decomposition of the Wilson loop for the fundamental representation of $G$ and the 't Hooft loop for the spinor representation of $G^{\vee}$ are $W=W_1+W_2$ and $H=H_1H_2$ respectively. Thus, we immediately conclude that $WH$ shows area law. On the other hand, because of the electric screening by the massive boson, $H=W_1H_1W_2H_2$ does not show area low. Therefore the charge for $H$ should be condensed in the vacuum. Under the $2\pi$ shift of theta angle of the underlying theory, we expect $(H_1,H_2)\to (W_1H_1, W_2H_2)$ since the index of embedding is one. From the above argument, in both vacua the charge $H$ is condensed thus we observe $H\to H$ under the shift. This is the expected behavior from a point of view of the Witten effect for $USp(2N)$ theory with $N=N_1+N_2=even$.

Finally we study the case $(r_1,r_2)=(W_1H_1,H_2), (H_1,W_2H_2)$. From the decomposition of the Wilson-'t Hooft loop
$${
WH=W_1H_1H_2+W_2H_1H_2,
}$$
it is easy to see that $WH$ does not show area law. Again these vacua have  $t=2$, so we conclude that the charge corresponding to $WH$ is condensed. Since these two vacua are connected by the $2\pi$ shift of the theta angle of $USp(2N)$ theory, the behavior $WH\to WH$ under the shift is what is expected from the argument in the Appendix.

%%%%%%%%%%%%%%%%%%%%%%%%%%%%%%%%%%%%
\subsection{$E_6 \to (SU(6)\times U(1))/\mathbb{Z}_2$}
%%%%%%%%%%%%%%%%%%%%%%%%%%%%%%%%%%%%

As the last example, we study the behavior of Wilson-'t Hooft loops in vacua with the confinement index $t=3$ for the theory discussed  in section \ref{E6II}. To determine which charges are condensed in the vacua, consider again decompositions of Wilson-'t Hooft loops $W^kH$, $k=0,1,2$. We denote the Wilson and 't Hooft loops for $SU(6)$ by $W_1$ and $H_1$. Taking into account $W_1^6=1$ and the electric screening by the massive boson $W_{\bf 20}=W_1^3=1$, we can write down the decompositions as follows:
\begin{eqnarray}
WH &\to &  (W_1 H_1)^{-1}+(W_1^4 H_1)^{-1}+\cdots , \nonumber \\
W^2H & \to & (W_1^2H_1)^{-1}+ (W^5_1H_1)^{-1} +\cdots , \nonumber \\
W^3H=H & \to & (W_1^3 H_1)^{-1}+(W^6_1H_1)^{-1} +\cdots . \nonumber
\end{eqnarray}
Thus, the vacua $r_1=s,s+3$, $s=0,1,2$, the charges associated with ${W^sH}$ are condensed. The other loops show area law. Since the index of embedding of $SU(6)$ is one\footnote{We use the branching rule for the fundamental representation of $E_6$ group. The index of ${\bf 27}$ representation in $E_6$ group is $\mu_{\bf 27}=6$. On the other hand the indices for ${\bf \bar{6}}$ and ${\bf 15}$ representations in $SU(6)$ group are $\mu_{\bf \bar{6}}=1$ and $\mu_{\bf 15}=4$ respectively. Thus from \eqref{indexofembedding}, we obtain the index of embedding for $SU(6)$ group $J={(1+1+4)/ 6}=1$.} under the $2\pi$ shift of the theta angle each loops transform as $H\to WH\to W^2H \to H$. This is what expected from the argument with weight vectors.

%%%%%%%%%%%%%%%%%%%%%%%%%
\section{'t Hooft's oblique confinement\label{oblique}}
%%%%%%%%%%%%%%%%%%%%%%%%%

The pure (ordinary, non-supersymmetric)  $SU(N)$ Yang-Mills theory was discussed by 't Hooft  \cite{tHooftConf}  under the assumption that  the system effectively reduces to an  Abelian $U(1)^{N-1}$ theory.  The relevant magnetic monopoles are those which manifest themselves as the singularities of the Abelian gauge fixing.  For $SU(2)$ case, 
 the particles are labeled by the electric and magnetic quantum numbers $(n_{m}, n_{e})$.  
   When $\theta$ is varied from $0$ to $2\pi$, a $(n_{m}, n_{e})$ particle  acquires
a electric charge by Witten's effect, 
  \[     Q = n_{e} +  \frac{\theta}{2\pi}  n_{m}.
  \]
Around $\theta=0,$    the $(1,0)$  magnetic monopole  is assume to condense and the system is in a confinement phase.  The confinement index is $t=2$. This is of course the well known dual superconductor picture of confinement in (an $SU(2)$ version of)  QCD.

He further  notes  that at $\theta=\pi$, 
the $(1,0)$  ``monopole''  and $(1,-1)$ ``dyon'' (in 't Hooft's phrasing, a composite of a monopole and gluon)    acquire electric charges  $\pm \frac{1}{2}$, respectively, by the Witten effect. They are strongly attracted to each other as in  the infrared the original $SU(2)\sim U(1)$ ``electric'' interactions  become very strong.  They therefore might form a tight bound state and condense.  In this  ``oblique confinement''  phase,  what condenses has the quantum numbers  $(n_{m}, n_{e})= (2,-1)$   but actually  it has only magnetic charge, no electric charge.  It is a pure magnetic monopole of charge $2$.  

Such a phase is not one of the   $(\mathbb{Z}_{N}, \mathbb{Z}_{N})$  classifications of possible phases of $SU(N)$ pure Yang-Mills, discussed in Donagi-Witten's paper \cite{DW}.  't Hooft's oblique confinement phase involves  strongly bound monopoles and gluons.   What is the confinement index of the theory in this phase?   't Hooft argues that in this phase the quarks (having the electric charge   $  \frac{1}{2}$),  can form an electrically neutral bound state with the $(1,-1)$ dyon  and escape confinement. Because of the peculiar (but well-known) property of the monopoe-quark composites,  the quarks escape confinement and travel to infinity as bosons.   The confinement index is $t=1$   in this case.

 An analogous phenomenon could in principle occur in the $SU(3)$ theory around $\theta= \frac{2 \pi}{3}$ \cite{tHooftConf}, where a magnetic monopole of charge $3$, which is a bound state of  monopoles and  gluons condenses. 

%It is interesting that in pure $SU(N)$  (non-supersymmetric) gauge theory such a rich variety of confinement (or Higgs) phase is possible in principle, depending on the dynamics, and the confinement index reflects some of these phenomena. 

%'t Hooft's  oblique confinement phase is realized in one of the vacua of the softly broken ${\cal N} =2$, $SU(2)$ theory with $N_{f}=3
%$ massless flavors \cite{SW,KonishiTerao}.

%%%%%%%%%%%%%%%%%%%%%%%%%%%%%%%%%%%%%%%%%%%%
\section{Softly broken ${\cal N} =2$ supersymmetric gauge theories \label{softly}}
%%%%%%%%%%%%%%%%%%%%%%%%%%%%%%%%%%%%%%%%%%%%

Exactly solved ${\cal N} =2$ theories (Seiberg-Witten solutions) with various gauge groups and matter contents
 teach us much  about confinement.   This is so because in these theories,  one often has an exact knowledge of the low-energy degrees of freedom and  of their dynamics.  When a soft ${\cal N} =1$  perturbation (the mass term   $\mu \, \Phi^{2}$ for the adjoint scalar superfield  $\Phi$)  is added to the system,  most of the degenerate vacua  of the ${\cal N} =2$  theories are lifted, and most of the surviving supersymmetric vacua are in a confinement phase,
 although this is not always the case  \footnote{For instance, in the special vacua at  the ``baryonic branch''  root  of  ${\cal N} =2$  $SU(N)$ SQCD,  or in  similar  ``special vacua''  present in large-flavored $SO(N)$ or $USp(2N)$  theories, the pertubation  $\mu \, \Phi^{2}$ does not induce confinement \cite{Konishione,Konishitwo}.}.  
 
If we restrict ourselves to the vacua of this type (i.e., in a confinement phase) only, there are roughly speaking four categories among the vacua in softly-broken ${\cal N} =2$ theories.  

\subsection{Abelian dual superconductor and nontrivial vacuum rearrangement\label{AbDual}} 

This is the famous case of $SU(2)$ ${\cal N} =2$ pure Yang-Mills theory, perturbed by the ${\cal N} =1$, $\mu \, \Phi^{2}$ term.  The perturbation lifts the vacuum degeneracy almost entirely, leaving only two vacua at $u= \pm \Lambda^{2}$, corresponding to the monopole/dyon singularities.  At each vacuum the monopole or dyon condenses and induces confinement \cite{SW}. This system is of course very  well known,  but we shall review it briefly nonetheless, as it illustrates some features of the condensation mechanism explicitly, and hence is a useful complement to the discussion of
Sections~\ref{confind} and \ref{condch}.

At  $u=\Lambda^{2}$, $\theta=0$ and the massless $(n_{m}, n_{e})=(1,0)$ monopole condenses,  and the system is in a confinement phase.  The magnetic $U(1)$ field is $A_{D}^{\mu}$, the dual of $A_{\mu}$.  The vacuum is in a dual superconducting state:  quarks would be confined by the chromoelectric vortex: the confinement index is $t=N(=2)$. 

At $u=-\Lambda^{2}$, the field  which is massless and condenses is a $(1,-1)$  ``dyon'' ,   or $(1,1)$ ``dyon'',  depending on how one  reaches  $u=-\Lambda^{2}$ from  $u=\Lambda^{2}$.  If one moves in the upper half $u$ plane one gets the   $(1,-1)$ particle becoming massless;  if the path is taken in the lower half plane the  $(1,1)$ particle becomes massless.  An important point is that the  $(1,-1)$  ``dyon'' at the   vacuum $u= e^{+i \pi } \Lambda^{2}$  is really a pure magnetic monopole.  Due to the Witten effect,  $\theta= 2  \,{\rm Arg} \, u =  2\pi$,  its electric charge  (defined by the coupling to the orginal gauge field $A_{\mu}$)  is 
\[     Q=  -1  +  \frac{\theta}{2\pi}\cdot 1 =0:
\]
it is a pure magnetic monopole, although the magnetic dual for this system is now defined by the gauge field   $A_{D}^{\mu}+ A^{\mu}$,  which is clearly non-local with respect to both $A^{\mu}$ and $A_{D}^{\mu}(x)$.

% In the limit $\mu \to \infty$  one reaches pure ${\cal N} =1$ $SU(2)$ Yang-Mills theory, and  the %result above  is clearly consistent with the statement about such theories mentioned already:   all  %$N$ different types of confining vacua occur once each in the $N$ vacua, with charges %condensed in the $k$-th vacuum being associated with the Wilson-'t Hooft loops  $W^k H$, %$k=0,\cdots N-1$.

As $u$ (or $\theta$) is varied  adiabatically the massless $(1,0)$ state  becomes massive at $u=-\Lambda^{2}$ (with mass $\sqrt{2} | a_{D}(u)|$).
As one approaches the  $u=e^{+i \pi } \Lambda^{2}$  vacuum, the  $(1,-1)$  ``dyon''  with mass $\sqrt{2} | a_{D}(u) - a(u)|$
which was massive at $u=\Lambda^{2}$  becomes instead massless  and condenses upon the ${\cal N} =1$ perturbation.  There is a nontrivial vacuum  rearrangement.  Even though the spectrum of the theory at  $u=\Lambda^{2}$ and that at $u=-\Lambda^{2}$  are identical,  in accordance with  an exact   discrete $\mathbb{Z}_{2}$ symmetry of the system, there is a nontrivial spectral flow. 

 Indeed, by further perturbing the system with a much smaller supersymmetry breaking parameter such as a nonvanishing  gluino mass,  it can be  shown explicitly \cite{Evans,KonishiSusy} that the system  makes a phase transition  from the first supersymmetric vacuum (where the $(1,0)$ field is condensed) to the second  (where $(1,-1)$ field is condensed) at $\theta=\pi$ \footnote{Here oblique confinement \`a la 't Hooft mentioned in Section~\ref{oblique} does not take place: the system is weakly coupled at all values of $\theta$.  When the bare $\theta$ parameter is varied from $0$ to $2\pi$,  the physical, low-energy $\theta_{phys}$ remains always small in magnitude, due to renormalization effect induced by instantons \cite{KonishiSusy}.   }. 

%Similarly, spectral flow and condensation mechanism of the $N$ distinct vacua of the  ${\cal N} =1$ $SU(N)$  Yang-Mills systems can be made  explicit,  by  softly breaking (by mass $\mu \Phi^{2}$)  pure ${\cal N} =2$  gauge theories and by using the known exact solutions \cite{SW,SWflavor}.
Similarly,  ${\cal N}=1$  supersymmetric (pure) Yang-Mills theories with $G= SU(N+1), SO(2N), $ $ SO(2N+1), \- USp(2N)$ gauge groups, can be studied by deforming the corresponding ${\cal N}=2$ pure YM theories by the adjoint scalar mass term $\mu \, \Phi^{2}$.  In all supersymmetric vacua the effective low-energy theory is a maximally Abelian $U(1)^{N}$ gauge theory,  with $N$ massless Abelian monopoles  \cite{DS,SWflavor,KonishiRicco}. The monopoles of various $U(1)$ charges  all condense simultaneously upon the adjoint mass perturbation \cite{AD,Konishione}.  Confinement in these theories is a dual Abelian superconductor of  't Hooft's type.

A word of caution is necessary, however.  The (weakly-coupled) Abelian dual superconductor picture as discussed here 
holds  only as long as $\mu \ll \Lambda$,  where $\Lambda$ is the scale of the ${\cal N}=2$ theories.  These theories dynamically Abelianize below $\Lambda$.  When $\mu \ge \Lambda$, the massive ``W'' bosons can be produced from the vacuum, connecting different $U(1)$ charges: at $\mu \to \infty$
(pure ${\cal N}=1$ Yang-Mills) the system is non-Abelian \cite{Strassler2}.

\subsection{Condensation of monopole-dyon composite: 't Hooft scenario \label{obliquebis}}

What happens in  one of the two vacua of the softly broken ${\cal N} =2$, $SU(2)$ theory with $N_{f}=3
$ {\it massless} flavors deserves a separate discussion.  This case nicely illustrates  the general fact that the gauge symmetry breaking
pattern (here $SU(2) \to U(1)$)  does not uniquely specify the  infrared dynamics of the system, i.e.,  which field is condensed and how global symmetry is broken, what the confinement index is, etc.,  just as in some ${\cal N}=1$ systems discussed in Section~\ref{confind}  and in Section~{\ref{condch}.

The massless ``dyon''  appearing in this vacuum (vacuum 1) has  $(2,1)$ 
 charges corresponding to Wilson-'t Hooft loop  $W H^{2}$,  but  it has vanishing physical electric charge due to the Witten effect, as $\theta = \pi$ there.  This massless particle is a singlet of the global chiral  $SO(6)$ symmetry group \cite{SW}.   

This is quite in contrast to the situation in the  other vacuum (vacuum 2), where the massless matter are  $(1,0)$  monopoles
carrying flavor  charge ${\bf 4}$ of  $SU(4)\sim Spin(6)$.  ${\cal N}=1$ perturbation induces condensation of these monopoles 
(confinement) and at the same time,  chiral symmetry breaking 
\[    SO(6)    \to    SU(3)\times U(1). 
\]
In contrast, no chiral symmetry breaking takes place in vacuum 1:  
neither the $(1,0)$ monopole nor $(1,1)$ dyon condenses. Due to the Witten effect, these latter particles carry
electric charges $\pm \tfrac{1}{2}$ and strongly attract each other.  The $(2,1)$ particle which becomes massless and condense in this vacuum can  naturally be identified \cite{KonishiTerao} as a composite of the $(1,0)$ monopole in ${\bf 4}$ of $SU(4)$ and the $(1,1)$ dyon in ${\bf  4}^*$ of $SU(4)$, 
forming  a singlet 
\[  {\bf 4}  \otimes {{\bf 4}^{*}}  = {\bf 1 } + \ldots.    \] 
Thus this system explictly realizes  't Hooft's oblique confinement phase.

\subsection{Confining vacua: non-Abelian dual superconductor \label{NAdual}}

In the  $r$-vacua of the softly  broken ${\cal N} =2$  $SU(N)$  supersymmetric QCD,  light non-Abelian monopoles   (massless monopoles carrying  non-Abelian dual  $SU(r)$ charges)  condense upon  ${\cal N}=1$ perturbation,  and induce confinement \cite{Konishione}. 
These are examples of confinement vacua of non-Abelian type:    dual superconductor of non-Abelian variety. 
As massless quarks are present in the theory, the confinement index is $t=1$.    

A fact that does not seem to be widely appreciated is that these {\it non-Abelian confining vacua}  occur very generally in this class of models. Among the  vacua of the softly broken ${\cal N} =2$ supersymmetric theories with $SU(N)$, $SO(N)$ and $USp(2N)$  gauge groups with quarks in the fundamental representation,  most are of this type.  
The  $r$-vacua of  ${\cal N} =2$  $SU(N)$ SQCD,  with  $r=2, 3, \ldots, N_{f}/2$ are all non-Abelian;  the same infrared  $r$-vacua appear in the  $SO(N)$ and $USp(2N)$ theories with nonvanishing equal bare quark masses \cite{Konishione,Konishitwo}.  Abelian dual superconducting vacua ($r=0,1)$ occur rather as an exception.

\subsection{Strongly-interacting magnetic monopole and dyon  composites}  

An intriguing  class of confining theories found in the context of softly-broken ${\cal N}=2 $ theories are the ``almost superconformal'' vacua  \cite{AGK,MKY}.   These confining vacua arise upon ${\cal N} =1$  perturbation of an ${\cal N} =2$ 
 nontrivial SCFT \cite{AD}, a system in which relatively non-local and strongly interacting  dyons and monopoles appear together in the low-energy theory. They are necessarily non-local theories:  they are  known in general as Argyres-Douglas vacua. 
 
   It is quite difficult to analyze these systems, especially when this SCFT has a non-Abelian gauge symmetry, 
  but a detailed study of the low-energy degrees of freedom and the known pattern of global symmetry breaking suggest that  
  confinement mechanism in these systems are markedly different from the cases of weakly coupled Abelian (Subsection~\ref{AbDual})  or non-Abelian dual superconductor (Subsection~\ref{NAdual}) vacua:   the condensed field  is not associated to a single monopole or dyon
  of lowest charges,  but to a composite of  monopoles and dyons,  a little like in  't Hooft's ``oblique confinement'' phase. These condensates induce confinement and at the same time break the global symmetry of the system \cite{MKY}. 
   
A wide class of ${\cal N}=2$ theories lead to confining vacua of this kind.  In fact, the $r=\tfrac{N_{f}}{2}$ vacua of $SU(N)$ theory as well as {\it all}  of the confining vacua of softly broken $SO(N)$ and $USp(2N)$  theories with {\it  vanishing bare quark masses} belong to this class of systems  \cite{Konishione,Konishitwo}.

%Although this story may not directly related to the main subject of  our paper,  please have a look at what I have studied a few years ago \cite{MKY}, with Yokoi san.    I'll put a remarks on these vacua too.

%%%%%%%%%%%%%%%%%%%%%%%%%%%%
\section{Discussion \label{concl}}
%%%%%%%%%%%%%%%%%%%%%%%%%%%%%

In this paper, we studied the behavior of the Wilson-'t Hooft loops for mainly ${\cal N}=1$ supersymmetric gauge theories. We computed the confinement index at the energy scale $E$ for various breaking patterns under the assumption $\Delta > E \gg \Lambda$. This condition is not essential but technically required. It allows us to treat the Higgsing semi-classically and to control nonabelian monopoles formed by the symmetry breaking. We could have reproduced the same results via a low-energy description by supposing $\Delta \gg \Lambda \simeq E$. Since the seminal paper by Cachazo, Douglas, Seiberg and Witten \cite{CDSW}, the low-energy descriptions of the theories have been studied intensively (see for example \cite{Review}) in terms of the generalized Konishi anomaly equations. For the $SU(N)$ gauge theory, the confinement index was reproduced from such perspective \cite{CSW}. However, as our calculations of the index suggest, their argument seems to be specific to the $SU(N)$ theory and is not necessarily applicable to other gauge theories: The dual coxeter number and the order of the center for the $SU(N)$ group are the same, thus the period integrals of a Reimann surface and a function on it, which are a solution of the generalized Konishi anomaly equations, could reproduce the index correctly. However in general they are not the same (See table 1). Therefore, it is not obvious how to extract the confinement index that we computed in this paper from such low-energy perspectives. Moreover the generalized Konishi anomaly equations for the gauge theories with $E_n$ gauge groups themselves are not known yet. Therefore it would be interesting to explore further on these avenues and clarify such physical properties for the low-energy description.

The similar issue can be seen in $SO/USp$ gauge theory with adjoint Higgs. In these models, as argued in the main text, the factorization of the Seiberg-Witten curve gives a powerful tool to analyze the low-energy behavior. As shown in \cite{AFO1}, there is a branch which can be understood by the multiplication map from the one for a theory with a lower-rank gauge group. For example for $SO(even)$ case, we can construct a map from a branch in $SO(2N)$ theory with a quartic superpotential to the corresponding one  in $SO(2KN-2K+2)$ theory with the same superpotential,
\begin{eqnarray}
P_{2KN-2K+2}=2\eta^K x^2 \Lambda^{2KN-2K} {\cal T}_K \left(P_{2N}\over 2\eta x^2 \Lambda^{2N-2} \right)
\end{eqnarray}
where $\eta$ is $2K$-th root of unity and ${\cal T}_K$ is the first kind Chebyshev polynomial. ÊContrary to the $SU(N)$ case, this number $K$ is not necessarily equal to the confinement index in that branch. Understanding the physical meaning of  $K$ in various gauge theories is an interesting issue,  which we will leave as an open problem.

\section*{Acknowledgments}

It is a pleasure to thank T.~Kawano for collaboration, for reading the manuscript carefully and giving us several useful suggestions. We would like to thank S.~Ashok, F.~Cachazo, J.~Gomis, K.~Intriligator, 
N. Saulina and Y. Tachikawa for useful comments and especially T.~Okuda for useful 
discussions and comments at various stages of this project. Research at Perimeter Institute for Theoretical Physics is supported in part by the Government of Canada 
through NSERC and by the Province of Ontario through MRI. 
YO are grateful to University of Pisa for kind hospitality and the organizers of Summer Institute 2009 at Fujiyoshida, Japan, where this work was partly carried out.

\bigskip

%%%%%%%%%%%%%%%%%%%%%%%%%%%%%%%%%%%%%%%%%%%%%%%%%%%%%%%%%%%%%%
\appendix

\setcounter{equation}{0}
\renewcommand{\theequation}{A.\arabic{equation}}
\appendix

%%%%%%%%%%%%%%%%%%%%%%%%%%%%
\section{{Vacua} in pure ${\cal N}=1$ Yang-Mills Theories}
%%%%%%%%%%%%%%%%%%%%%%%%%%%%

We would like to know how the Wilson-'t Hooft loops behave in each vacuum of ${\cal N}=1$ pure gauge theories. According to GNO \cite{GNOW}, electric and magnetic charges of $G$ are represented by weight vectors $(\mu, \nu)$ of $G$ and its dual $G^{\vee}$,  respectively. With this representation, the quantization condition of charge is written as $2 \mu\cdot \nu \in \mathbb{Z}$. A Wilson-'t Hooft loop corresponding to $(\mu, \nu)$ is labelled by the center and the fundamental group $\pi_1$. For example, in $SU(N)$ pure Yang-Mills theory they are $\mathbb{Z}_N\times \mathbb{Z}_N$. 

Now consider the Witten effect for the weight vectors. Since each massive vacuum in pure Yang-Mills theory are connected by the $2\pi $ shift of theta angle, it is useful to know the Witten effect for the weight vectors itself to understand the behavior of  Wilson-'t Hooft loops in each vacuum. An interesting paper \cite{Anton1} discussed S-duality for the weight vectors in a ${\cal N}=4$ theory. However as far as the shift of theta angle is concerned, supersymmetry does not really matter. The formula shown in the paper is applicable to ${\cal N}=1$ theories as well. According to \cite{Anton1}, under the $2\pi $ shift, the weight vectors transform as 
$${
(\mu, \nu)\to (\mu+\nu^*, \nu),
}$$
where $\nu^*$ means as follows: The $\nu$ is a weight vector in the dual gauge group which is a coweight vector of the electric group while $\nu^*$ should be an electric weight. To begin with, let us start with a root vector of the electric group. A coroot is defined by $\alpha^{\vee} ={2 \alpha / (\alpha, \alpha)}$. If the gauge group is $ADE$, then this coroot and root are equivalent. In a convention where the Cartan metric is normalized such that $(\alpha, \alpha)=2$, for example see \cite{GW},
$\alpha^{\vee} =\alpha$ for $ADE$. On the other hand, for the non-simply raced cases, $BCFG$, there are two kinds of roots, $\alpha^{\vee} = \alpha$ for short root, $\alpha^{\vee}= \alpha/ n_g$ for long root where $n_g=2$ for $BDF$ and $n_g=3$ for $G_2$. Following the argument shown in Appendix in \cite{GW}, the $*$ operation for the root vectors is given by  
\begin{eqnarray}
(\alpha^{\rm long})^*= {1\over n_g} \alpha^{\rm long}=\tilde{\alpha}^{\rm short},\quad (\alpha^{\rm short})^*=\alpha^{\rm short}=\tilde{\alpha}^{\rm long}, \label{star}
\end{eqnarray}
where tilde means the root vectors in dual group.
With this in mind, let us consider the Witten effect for various groups.

\vspace{0.2cm}
%%%%%%%%%%%%%%%%%%%%%%%%%%%%%%%%%%
$\bullet$ $Spin(2N+1)$ Yang-Mills
%%%%%%%%%%%%%%%%%%%%%%%%%%%%%%%%%%

\noindent
%:
If all the matter contents are trivial under the center, then global structure of the group may be $H=Spin(2N+1)/\mathbb{Z}_2$. Since the fundamental group is nontrivial $\pi_1(H)=\mathbb{Z}_2$, there are monopoles in this theory, which  belong to representations in $H^{\vee}=USp(2N)$ group. One of the non-trivial magnetic weight vectors is given by the highest weight of fundamental representation of $USp(2N)$ group. The weight vector can be written in terms of the simple roots with rational number coefficients because the weight vector does not belongs to the root lattice,
$${
\nu= \sum_{r=1}^{N-1}  \alpha^{(r)} +{1\over 2}\alpha^{(N)}.
}$$
Using the \eqref{star}, we can get $\nu^*$ as follows:
\begin{equation}
 \nu^* = \sum_{k=1}^{N-1} \tilde{\alpha}^{(k)} +\tilde{\alpha}^{ (N)},\label{starUSp}
\end{equation}
where we used the fact that $\alpha^{(N)}$ is a long root and others $\alpha^{(r)}$ are short root. The \eqref{starUSp} clearly shows that $\nu^*$ is an element of the root lattice of $Spin(2N+1)$ group. So a Wilson loop corresponding to the charge $\nu^*$ is trivial. Therefore under the $2\pi $ shift of the theta angle 't Hooft loop $H$ for the fundamental representation of $USp(2N)$ transforms into the same $H$. Thus we conclude that all $2N-1$ vacua in $Spin(2N+1)$ theory have the same behavior in the Wilson-'t Hooft loops. In particular, we claim that all vacua are in confinement phase: none is in  an oblique confinement phase.  Since the original pure Yang-Mills theory does not include an electric charge for the spinor representation and the Witten effect also does not generate such charge, there is no reason to believe a condensation of a dyon carrying such electric charge without assuming a nontrivial dynamics at low energies. This argument is consistent with the result shown in \cite{WittenIndex}.

\vspace{0.2cm}
%%%%%%%%%%%%%%%%%%%%%%%
$\bullet$ $USp(2N)$ Yang-Mills
%%%%%%%%%%%%%%%%%%%%%%%%

\noindent
We consider the pure ${\cal N}=1$ $USp(2N)$ gauge theory in a similar fashion. Again the matter fields are all trivial under the center, the global structure of the group is $H=USp(2N)/\mathbb{Z}_2$. Since the fundamental group is nontrivial,  $\pi_1(H)=\mathbb{Z}_2$, there is a monopole  whose representation is in the dual group, $H^{\vee}=Spin(2N+1)$. The most elementary representation of the dual group is the spinor representation. The highest weight of the spinor representation can be written in terms of simple roots of the dual group,
$${
\nu= {1\over 2} \big(  \sum_{k=1}^{N} k \alpha^{(k)} \big).
}$$
Using the relation \eqref{star}, we get $\nu^*$ as follows:
$${
 \nu^* = \sum_{i=1}^{N-1} \tilde{\alpha}^{(i)}+{N\over 2}\tilde{\alpha}^{(N)},
}$$
where we used that the fact that the first $N-1$ simple roots $\alpha^{(r)}$ are long roots and $\alpha^{(N)}$ is a short. 
From this, we see that if $N$ is even, then the $\nu^*$ is in the root lattice of the original group $USp(2N)$. On the other hand, if $N$ is odd then it does not belong to the root lattice. This means that behavior of 't Hooft loop for the monopole under the $2\pi$ shift of theta angle is different. For $N={\rm even}$, $H$ stays in the same class while for $N=$odd, $H$ goes to different class $WH$. By the same reasoning as  in the previous example, we conclude that for $N={\rm even}$ all vacua are in the confinement phase. Again this is consistent with the argument by Witten \cite{WittenIndex}.

For simply-raced groups, the behavior of the  't Hooft loop corresponding to the fundamental representation of the dual groups is the same for all cases except $Spin(4k)$ theory. By the $2\pi$ shift of the theta angle, $H$ goes to $WH$. For the $Spin(4k)$ theory, electric and magnetic charges are labelled by $(\mathbb{Z}_2\times \mathbb{Z}_2)_{\rm ele}\times (\mathbb{Z}_2\times \mathbb{Z}_2)_{\rm mag}$. Following the argument by 't Hooft \cite{tHooft}, finding a massive vacua is choosing a $\mathbb{Z}_2\times \mathbb{Z}_2$ subgroup. In $Spin(4k)$ theory there are four types of vacua generated by the following charges,
$${
\{H, H^{\prime} \},\quad \{WH ,W^{\prime} H^{\prime} \},\quad \{ H, W^{\prime}H^{\prime} \},\quad \{WH, W^{\prime} \}  .
}$$
Here again, we expect that vacua include at least one confinement phase. The first two are transformed into each other under the $2\pi$ shift of the theta angle thus these two are realized in vacua for the $Spin(4k)$ theory.

%%%%%%%%%%%%%%%%%%%%%%%%%%%%%%%%%%%%%%%%%%%%%%
%
% Bibliography
%
%%%%%%%%%%%%%%%%%%%%%%%%%%%%%%%%%%%%%%%%%%%%%%

\end{document}